\documentclass[12pt]{iopart}
\usepackage{iopams}
\usepackage{graphicx}
\newcommand{\ket}[1]{\left|#1\right\rangle} %kets
\newcommand{\bra}[1]{\langle#1|} %bras
\newcommand{\hf}{\ensuremath{\frac{1}{2}}}
\newcommand{\tv}{\ensuremath{\frac{3}{2}}}

\newcommand{\mod}{\textrm{ mod }}
\newcommand{\F}{\mathcal{F}}
\newcommand{\Po}{\ensuremath{P_{1,2}}}
\newcommand{\Pt}{\ensuremath{P_{2,3}}}

\begin{document}

\title{Exchange-only dynamical decoupling in the 3-qubit decoherence
  free subsystem}
\author{J R West and B H Fong}
\address{HRL Laboratories, LLC, 3011 Malibu Canyon Road, Malibu, CA,
  90265, USA}
\ead{bhfong@hrl.com}

\begin{abstract}
  The Uhrig dynamical decoupling sequence achieves high-order
  decoupling of a single system qubit from its dephasing bath through
  the use of bang-bang Pauli pulses at appropriately timed intervals.
  High-order decoupling of single and multiple qubit systems from
  baths causing both dephasing and relaxation can also be achieved
  through the nested application of Uhrig sequences, again using
  single-qubit Pauli pulses.  For the 3-qubit decoherence free
  subsystem (DFS) and related subsystem encodings, Pauli pulses are
  not naturally available operations; instead, exchange interactions
  provide all required encoded operations.  Here we demonstrate that
  exchange interactions alone can achieve high-order decoupling
  against general noise in the 3-qubit DFS.  We present decoupling
  sequences for a 3-qubit DFS coupled to classical and quantum baths
  and evaluate the performance of the sequences through numerical
  simulations.
\end{abstract}

\section{Introduction}
Dynamical decoupling (DD) pulse sequences have had a long history
beginning with the Hahn spin echo \cite{Hahn:1950fk} and the
Carr-Purcell-Meiboom-Gill sequences \cite{Carr:1954uq,Meiboom:1958kx}
and continuing to the present.  Recently, Uhrig has developed a DD
sequence that, by varying the time intervals between pulses, is able
to decouple transverse dephasing to order $n$ in the system-bath
coupling strength with $n+1$ pulse intervals
\cite{Uhrig:2007kx,Uhrig:2008vn}.  The Uhrig decoupling sequence is in
fact universal, decoupling both classical and quantum baths that cause
either transverse dephasing or longitudinal relaxation
\cite{Yang:2008uq}.  Several generalizations of the Uhrig sequence,
and its use of non-uniform pulse intervals, have now been made.  These
generalizations allow high-order decoupling of a single qubit from
dephasing noise baths with different noise power spectra
\cite{Biercuk:2009ly,Uys:2009qf}, decoupling of a single qubit from
baths causing simultaneous dephasing and relaxation
\cite{West:2010oz}, and decoupling of multi-qubit systems from general
baths \cite{Wang:2011dq}.  The Uhrig decoupling scheme and its
generalizations all require single qubit Pauli pulses for
implementation.

In semiconductor quantum dot systems, however, single qubit operations
are not easily implemented, in contrast with the two qubit exchange
interaction.  While the exchange interaction can be performed in
sub-nanosecond time scales \cite{Petta:2005vn}, single qubit rotations
may be two orders of magnitude slower or more
\cite{Koppens:2006vn,Nowack:2007ys}, and can be technically more
demanding.  The difference in requirements between single and two
qubit gates has led to the development of encodings that use the
exchange interaction alone
\cite{Bacon:2000qf,Kempe:2001a,Kempe:2001b}.  The smallest such
encoding is the 3-qubit decoherence free subsystem (DFS), for which
explicit exchange gate sequences for encoded universal computation are
given in \cite{DiVincenzo:2000,Fong:2011fk}.  Creation of a DFS itself
may also be performed using exchange pulses alone \cite{Wu:2002vn}.
While encoded computation can be performed with exchange gates alone,
the use of the DD pulse sequences described above would require single
qubit Pauli operations.  Here we demonstrate that exchange gates alone
suffice for high-order decoupling of the DFS-encoded information from
general baths.

Our new exchange-only DD sequences explicitly take advantage of the
decoherence free properties of DFS encodings.  Because the 3-qubit DFS
is decoherence free with respect to any global interaction, decoupling
from a decohering bath can be achieved by globalizing any local
interactions to high order.  In contrast to standard Pauli-based
decoupling schemes, our aim is not to cancel system-bath coupling
terms, but to globalize them, so that the effective Hamiltonian
created by the DD scheme causes only the gauge qubit and bath to
evolve.  This idea applies to any subsystem encoding: decoupling need
only preserve the subsystem of interest while gauge subsystems can
evolve arbitrarily.  An alternative method for low-order leakage
elimination using the simultaneous operation of multiple exchange
gates has been given in \cite{Wu:2002zr,Byrd:2005ys}.

The paper is organized as follows.  In section \ref{sec:background} we
briefly review aspects of the Uhrig dynamical decoupling sequence and
the 3-qubit decoherence free subsystem.  Section
\ref{sec:classical_dd} describes how exchange-only decoupling is
achieved for a 3-qubit DFS subject to dephasing from a classical noise
bath.  Section \ref{sec:quantum_dd} gives the analogous exposition for
decoupling from a quantum bath.  Simulations of the decoupling
sequences for a DFS qubit coupled to classical and quantum baths are
presented in section \ref{sec:simulations}.  We conclude in section
\ref{sec:conclusion}.  In the following we use the acronym DFS to
refer to decoherence free subsystem (rather than subspace).  We assume
``bang-bang'' exchange pulses perform the decoupling, i.e., the pulses
perform a perfect finite operation using infinite power in
infinitesimal time.

\section{Background}
\label{sec:background}

\subsection{Uhrig dynamical decoupling}
\label{sec:udd}
The Uhrig dynamical decoupling (UDD) sequence decouples a single qubit
from a classical dephasing bath \cite{Uhrig:2007kx,Uhrig:2008vn}.  The
Hamiltonian for such a system is given by
\begin{equation}
H=Z B(t),
\label{eq:ham}
\end{equation}
where $Z$ is the Pauli $Z$ operator and $B(t)$ is a time-dependent
real-valued function.  Subjected to a sequence of $\pi$ pulses about the
$x$-axis, the Hamiltonian in the toggling frame becomes
\begin{equation}
H=f(t) Z B(t),
\label{eq:hamt}
\end{equation}
where $f(t)$ is the UDD ``switching function'' and takes values of
$\pm1$.  The switching times, when $f$ switches between 1 and -1 (or
vice versa), give the instances when $\pi$ pulses are applied.  The
propagator corresponding to free evolution interspersed with $\pi$
pulses is given by
\begin{equation}
U(T)=\exp(-\rmi Z \theta(T)),
\label{eq:Ut}
\end{equation}
where $T$ is the total time of the DD sequence, and the accumulated
phase $\theta$ from the classical noise bath over time $T$ is 
\begin{equation}
\theta(T)=\int_0^T \rmd s f(s) B(s).
\label{eq:theta}
\end{equation}
For order $n$ decoupling, we require that 
\begin{equation}
0 = \int_0^T \rmd s f(s) s^p
\label{eq:constraint}
\end{equation}
for all $0\leq p<n$, which leaves only terms of order $T^{n+1}$ and
higher in the accumulated phase.  In other words, the DD sequence has
resulted in an effective Hamiltonian with system-bath coupling only at
order $T^{n+1}$ and higher.  Uhrig showed that the switching times for
order $n$ decoupling are given by the simple formula
\begin{equation}
t_j = T \sin^2\left(\frac{j\pi}{2(n+1)} \right),
\label{eq:UDD_switching_times}
\end{equation}
for $j=1,\ldots,n$.  It was subsequently shown that the UDD sequence
is in fact universal \cite{Yang:2008uq}, i.e., UDD decouples dephasing
due to both classical and quantum baths.

The propagator (\ref{eq:Ut}) evolves an initial state $\ket{\psi_i}$
to the final state $\ket{\psi_f}=U(T)\ket{\psi_i}$.  The fidelity of
memory preservation due to the decoupled evolution is then
\begin{equation}
F(T)=|\langle\psi_f|\psi_i\rangle|^2 = \cos^2\theta(T) +
\langle\psi_i|Z|\psi_i\rangle^2 \sin^2\theta(T).
\label{eq:fidelity}
\end{equation}
For an initial state on the $y$-axis of the Bloch sphere
$\langle\psi_i|Z|\psi_i\rangle=0$, and the performance of the DD
sequence can be described in terms of the decoherence function
$W(T)=\langle Y(T) \rangle$ \cite{Cywinski:2008ys}, which is the
expectation value of the Pauli $Y$ operator.  When
$\langle\psi_i|Z|\psi_i\rangle=0$, the fidelity and the decoherence
function are related through $F(T) = \frac{1}{2}(1+W(T))$.  For a
noise bath with Gaussian statistics, the ensemble averaged decoherence
function is
\begin{eqnarray}
  W(T) &\equiv& \rme^{-\chi(T)}, \label{eq:coherence}\\
  \chi(T) &=& \int_0^\infty \frac{\rmd \omega}{2\pi} S(\omega)
  |\hat{f}^{(n)}(\omega T)|^2.\label{eq:chi}
\end{eqnarray}
Here $S(\omega)$ is the bath's noise spectral density and
$\hat{f}^{(n)}(\omega T)$ is the Fourier transform of the UDD
switching function $f^{(n)}(t)$ of order $n$:
\begin{equation}
  \omega^2 |\hat{f}^{(n)}(\omega T)|^2 = 
\left| \sum_{k=1}^{n+1} f^{(n)}\left(\frac{t_{k-1}+t_k}{2}\right) 
\left(\rme^{\rmi \omega t_k} - \rme^{\rmi \omega t_{k-1}}\right)
\right|^2.
\label{eq:filter_function}
\end{equation}
Since $\{t_j\}$ are the UDD switching times, $(t_{k-1}+t_k)/2$ is the
midpoint of the $k^\mathrm{th}$ interval, where $f^{(n)}$ is
$(-1)^{k-1}$.  The expression in (\ref{eq:filter_function}) is the
``filter function''; together with (\ref{eq:coherence}) and
(\ref{eq:chi}) the filter function gives an interpretation of the UDD
sequence as a solution to a filter design problem
\cite{Biercuk:2011zr}.  For the UDD sequence, the filter function
gives a high-pass filter, suitable for decoupling noise spectra with a
sharp high-frequency cutoff.  The switching times can be adjusted to
decouple noise spectra with other frequency characteristics
\cite{Biercuk:2009ly,Uys:2009qf}.

\subsection{The 3-qubit decoherence free subsystem}
\label{sec:dfs}
The general theory of decoherence free subspaces and subsystems and
encoded universality is given in \cite{Kempe:2001a}.  Initialization,
measurement, and universal computation in the 3-qubit DFS are
described in \cite{DiVincenzo:2000,Fong:2011fk}.  In this subsection
we give a brief summary of the 3-qubit DFS to fix notation and to
provide the physical motivation behind the pulse sequence designs in
the following sections.

The states of three spin-$\hf$ qubits can be described by the quantum
numbers of three commuting operators $S$, $S_{1,2}$, and $S_z$. $S$ is
the total spin of all three qubits and distinguishes valid and leaked
states in the DFS encoding.  $S_{1,2}$ is the total spin of the first
two of the three qubits and determines the encoded qubit state.  $S_z$
is the total $z$-axis spin of the three qubits and gives the gauge
state.  Explicit definitions of these operators are given in
\cite{Fong:2011fk}. The states of the eight dimensional Hilbert space
of the three qubits are explicitly:
\begin{eqnarray}
 \ket{1}&=\ket{\hf,0,\hf}&=\frac{1}{\sqrt{2}}(\ket{010}-\ket{100})\label{eq:ket1}\\
 \ket{2}&=\ket{\hf,0,-\hf}&=\frac{1}{\sqrt{2}}(\ket{011}-\ket{101})\\
 \ket{3}&=\ket{\hf,1,\hf}&=\sqrt{\frac{2}{3}}\ket{001}-\frac{1}{\sqrt{6}}\ket{010}
   -\frac{1}{\sqrt{6}}\ket{100}\\
 \ket{4}&=\ket{\hf,1,-\hf}&=\frac{1}{\sqrt{6}}\ket{011}+\frac{1}{\sqrt{6}}\ket{101}-\sqrt{\frac{2}{3}}\ket{110}\label{eq:ket4}\\
 \ket{5}&=\ket{\tv,1,\tv}&=\ket{000}\\
 \ket{6}&=\ket{\tv,1,\hf}&=\frac{1}{\sqrt{3}}(\ket{001}+\ket{010}+\ket{100})\\
 \ket{7}&=\ket{\tv,1,-\hf}&=\frac{1}{\sqrt{3}}(\ket{011}+\ket{101}+\ket{110})\\
 \ket{8}&=\ket{\tv,1,-\tv}&=\ket{111}\label{eq:ket8}
 \end{eqnarray}
where states on the left in (\ref{eq:ket1})--(\ref{eq:ket8}) are assigned
labels 1--8, states in the middle are described in the angular
momentum basis $\ket{S,S_{1,2},S_z}$, and states on the right are
written in terms of the standard computational basis.

States $\ket{1}$--$\ket{4}$ in (\ref{eq:ket1})--(\ref{eq:ket4}) span
the valid subspace of the 3-qubit DFS.  States $\ket{1}$ and $\ket{2}$
are encoded 0 states, with gauge states $+\frac{1}{2}$ and
$-\frac{1}{2}$, respectively; states $\ket{3}$ and $\ket{4}$ are
encoded 1 states, with gauge states $+\frac{1}{2}$ and $-\frac{1}{2}$,
respectively.  In the valid subspace, the encoded quantum number
$S_{1,2}$ and the gauge quantum number $S_z$ may be considered as the
computational basis states of two two-state subsystems---two effective
qubits.  The states have been ordered so that the first subsystem
effective qubit gives the encoded state in the DFS and the second
subsystem effective qubit gives the gauge state.  Valid DFS states
must not only be in the valid subspace, but additionally be
factorizable between the encoded and gauge quantum numbers.  We also
define a projector onto the valid subspace
\begin{equation}
  \Pi \equiv \ket{1}\bra{1}+\ket{2}\bra{2}+\ket{3}\bra{3}+\ket{4}\bra{4},
\label{eq:projector}
\end{equation}
for use in section \ref{sec:classical_dd}. 

Encoded operations are performed with exchange gates between pairs of
constituent physical qubits.  Because exchange commutes with the total
angular momentum operators $S$ and $S_z$, exchange can only change the
encoded DFS quantum number while leaving the gauge and leaked quantum
numbers unaffected.  Similarly, any interaction comprised of total
spin component operators can only change the gauge quantum number
$S_z$ and leaves the encoded $S_{1,2}$ and leaked $S$ quantum numbers
unchanged.  The encoded qubit is thus decoherence free with respect to
any global interaction.  For a given system-bath coupling of the DFS
constituent physical qubits, our goal is to design a pulse sequence
resulting in an effective Hamiltonian whose system-bath interactions
to order $n$ contain only total spin component $S_x$, $S_y$, or $S_z$
operators.

\section{Decoupling the 3-qubit DFS from classical phase noise}
\label{sec:classical_dd}

Consider the following Hamiltonian coupling a 3-qubit DFS to a
classical constant dephasing noise bath:
\begin{equation}
H = H_1 = Z_1 B_1 + Z_2 B_2 + Z_3 B_3,
\label{eq:H_1}
\end{equation}
where $Z_j$ gives the Pauli $Z$ operator for physical qubit $j$, and
$B_j$ is a constant (in time) real number giving the bath strength at
qubit $j$.  Free evolution corresponding to (\ref{eq:H_1}) is $U_f(t)
= \exp(-\rmi H t)$.  We define the pulse $P$ as
\begin{equation}
P \equiv \Pt.\Po
\end{equation}
where $P_{i,j}$ is a full swap operation between qubits
$i$ and $j$.  The Hamiltonian in (\ref{eq:H_1}) conjugated singly and
doubly by the pulse $P$ gives the permuted Hamiltonians
\begin{eqnarray}
H_2 \equiv P^{-1}.H_1.P = Z_2 B_1 + Z_3 B_2 + Z_1 B_3 \label{eq:H_2},\\
H_3 \equiv P^{-1}.H_2.P = Z_3 B_1 + Z_1 B_2 + Z_2 B_3 \label{eq:H_3}.
\end{eqnarray}
Conjugation by $P$ causes the physical qubits to dephase under the
influence of the different local noise baths.  By spending equal time
in each of the three permutations in (\ref{eq:H_1}), (\ref{eq:H_2}),
and (\ref{eq:H_3})---the even permutations (alternating group $A_3$)
of the symmetric group $S_3$ on three elements---we generate an
effective global system-bath interaction.  The propagator
corresponding to equal time evolution in each permutation is given by
\begin{eqnarray}
U &=& \exp(-\rmi H_3 \tau).\exp(-\rmi H_2 \tau).\exp(-\rmi H_1 \tau) \label{eq:first_order_evolution}\\
 &=& \exp(-\rmi (Z_1 + Z_2 + Z_3)(B_1 + B_2 + B_3)\tau
 ) \label{eq:first_order_effective_Hamiltonian} \\
 &=& P.U_f(\tau).P.U_f(\tau).P.U_f(\tau). \label{eq:first_order_pulsing}
\end{eqnarray}
The effective Hamiltonian in
(\ref{eq:first_order_effective_Hamiltonian}) consists of the total
spin-$z$ operator alone and evolves only the gauge qubit.  We have
thus decoupled the encoded subsystem state from the dephasing bath to
first order.  The pulse sequence for this first order exchange-only
scheme is shown explicitly in (\ref{eq:first_order_pulsing}) and
involves free evolution and full swap (exchange) operations alone.  Of
course the bath functions are not generally constant, and in the
following we describe how this ``globalization'' of the Hamiltonian
can be accomplished to high order for general time-varying bath
functions.  We will see that free evolution for non-uniform time
intervals, interspersed with full swap operations, is sufficient for
high order decoupling.

For local classical dephasing baths with arbitrary time dependence the
Hamiltonian is
\begin{equation}
H = Z_1 B_1(t) + Z_2 B_2(t) + Z_3 B_3(t).
\label{eq:H}
\end{equation}
Since the effect of conjugation by $P$ or $P^{-1}$ is to permute the
local bath seen by the constituent qubits, the Hamiltonian in the
toggling frame (cf. (\ref{eq:hamt})) is
\begin{equation}
H = Z_1 B_{\alpha_1(t)}(t) + Z_2 B_{\alpha_2(t)}(t) + Z_3 B_{\alpha_3(t)}(t),
\label{eq:Ht}
\end{equation}
where $\alpha_j(t) = 1,2, \mathrm{or\ } 3$ and identifies which local
bath constituent qubit $j$ is experiencing.  Restriction to $A_3$
permutations completely specifies $\alpha_2(t)$ and $\alpha_3(t)$ in
terms of $\alpha_1(t)$: $\alpha_2(t) = \alpha_1(t) + 1 (\!\mod 3)$ and
$\alpha_3(t) = \alpha_1(t) + 2 (\!\mod 3)$, where the modulus is taken
with offset 1.  For notational convenience we use all three
$\alpha_j(t)$'s in the following.  The three types of Hamiltonians
$H_1$ (\ref{eq:H_1}), $H_2$ (\ref{eq:H_2}), and $H_3$ (\ref{eq:H_3})
(generalized to time-dependent local baths) have $\alpha_1$ values of
1, 3, and 2, respectively.

The $\alpha_j(t)$'s change values when $P$ or $P^{-1}$ pulses are
applied.  Between pulses the $\alpha_j(t)$ values are constant.  The
instances at which the pulses are applied are the switching times for
exchange-only decoupling.  Unlike the UDD situation, both the
switching times \textit{and} the Hamiltonian sequence (or
$\alpha_1(t)$) must be determined.  The UDD case involves only two
Hamiltonian types and one pulse type: the Pauli $X$ pulse toggles the
evolution back and forth between the two Hamiltonian types.  In the
DFS case, we may choose to apply either a $P$ or $P^{-1}$ pulse, i.e.,
after evolution under one of the Hamiltonians (\ref{eq:H_1}),
(\ref{eq:H_2}), or (\ref{eq:H_3}), the next evolution interval can be
chosen from either of the remaining two Hamiltonians.

The propagator associated with the modulated Hamiltonian (\ref{eq:Ht})
is 
\begin{equation}
U(T) = \exp\left(-\rmi Z_1 \theta_1(T)\right) 
\exp\left(-\rmi Z_2 \theta_2(T)\right)
\exp\left(-\rmi Z_3 \theta_3(T)\right),
\label{eq:U}
\end{equation}
where $\theta_j(T)$ gives the phase accumulated by constituent qubit
$j$ in time $T$,
\begin{equation}
\theta_j(T) = \int_0^T \rmd s B_{\alpha_j(s)}(s).
\label{eq:theta_j}
\end{equation}
If all the $\theta_j(T)$ values are equal, the propagator will contain
only a global interaction term generated by the total spin $S_z$,
which drives gauge evolution only.  Computing the fidelity of memory
preservation shows explicitly that setting all the $\theta_j(T)$ equal
leaves the encoded subsystem unchanged.  Given an initial DFS state
$\ket{\psi_e}\ket{\psi_g}$, where $\ket{\psi_e}$ gives the encoded
subsystem state and $\ket{\psi_g}$ gives the gauge subsystem state,
the fidelity of memory preservation under the evolution in
(\ref{eq:U}) is 
\begin{equation}
  F(T) = \sum_\mu \left| \bra{\mu} \bra{\psi_e}\Pi U(T) \Pi
    \ket{\psi_e}\ket{\psi_g}\right|^2.
\label{eq:fidelity3}
\end{equation}
The sum over $\mu$ is a partial trace over the gauge subsystem, since
we are concerned only with encoded subsystem preservation.  $\Pi$ is
the projector into the valid subspace defined in (\ref{eq:projector}).
Substituting the propagator (\ref{eq:U}) into the fidelity
(\ref{eq:fidelity3}) we find
\begin{eqnarray}
F &=& c_0 + c_1 \cos 2(\theta_2(T) - \theta_3(T)) + c_2 \cos
2(\theta_3(T) - \theta_1(T)) \nonumber\\
&&+ c_3 \cos 2(\theta_1(T) - \theta_2(T)),
\label{eq:fidelity3_explicit}
\end{eqnarray}
where the $c_j$'s depend on the initial encoded state $\ket{\psi_e}$
(and not the initial gauge state $\ket{\psi_g}$) and are given in
the Appendix; the $c_j$'s sum to 1.  

Equation (\ref{eq:fidelity3_explicit}) shows that setting all the
accumulated phases equal preserves the encoded subsystem fidelity.
From (\ref{eq:theta_j}) the accumulated phase difference
$\theta_1(T)-\theta_2(T)$ is
\begin{eqnarray}
\theta_1(T) - \theta_2(T) &=& \int_0^T \rmd s \left(
B_{\alpha_1(s)}(s) - B_{\alpha_2(s)}(s)\right) \label{eq:diff1_step}\\
& = &\int_0^T \rmd s \left(
f_1(s) B_1(s) + f_2(s) B_2(s) + f_3(s) B_3(s) \right) \label{eq:diff1}.
\end{eqnarray}
The switching functions $f_j(t)$ depend on the Hamiltonian type in
each evolution interval.  For evolution under $H_1$, qubit 1 sees bath
$B_1$ and qubit 2 sees bath $B_2$ so $\alpha_1=1$ and $\alpha_2=2$.
The phase difference between qubit 1 and qubit 2 gives a positive sign
on $B_1$, a negative sign on $B_2$, and no contribution from $B_3$ so
that $f_1=1$, $f_2=-1$ and $f_3=0$.  The switching functions and
$\alpha_j$ values for the other Hamiltonian types may be similarly
determined and are given in table \ref{tab:switching_functions}.  The
two remaining phase differences are
\begin{eqnarray}
\theta_2(T) - \theta_3(T) &=& \int_0^T \rmd s \left(
f_3(s) B_1(s) + f_1(s) B_2(s) + f_2(s) B_3(s) \right) \label{eq:diff2},\\
\theta_3(T) - \theta_1(T) &=& \int_0^T \rmd s \left(
f_2(s) B_1(s) + f_3(s) B_2(s) + f_1(s) B_3(s) \right) \label{eq:diff3}.
\end{eqnarray}
The same switching functions $f_j(t)$ appear here as in
(\ref{eq:diff1}).

\begin{table}
\caption{\label{tab:switching_functions} Bath identification functions
  $\alpha_j$ and switching functions $f_k$.}
\begin{indented}
\item[]\begin{tabular}{@{}cllllll}
\br
Hamiltonian type & $\alpha_1$ & $\alpha_2$ & $\alpha_3$ & $f_1$ & $f_2$ & $f_3$\\
\mr
$H_1$ & 1 & 2 & 3 & 1 &-1 & 0 \\
$H_2$ & 3 & 1 & 2 &-1 & 0 & 1 \\
$H_3$ & 2 & 3 & 1 & 0 & 1 &-1 \\
\br
\end{tabular}
\end{indented}
\end{table}
 
Decoupling of the DFS from arbitrary time-dependent classical baths
can be achieved by setting each term in (\ref{eq:diff1}),
(\ref{eq:diff2}), and (\ref{eq:diff3}) to zero, order-by-order.  For
order $n$ decoupling this results in the constraints
\begin{eqnarray}
0 &=& \int_0^T \rmd s f_1(s) s^p \label{eq:constraint1},\\
0 &=& \int_0^T \rmd s f_2(s) s^p \label{eq:constraint2},
\end{eqnarray}
for $0\leq p < n$ and a suitably chosen sequence of Hamiltonian types.
Because $f_3 = -(f_1+f_2)$, satisfying (\ref{eq:constraint1}) and
(\ref{eq:constraint2}) automatically constrains $\int_0^T \rmd s
f_3(s) s^p$ to be zero.  Equations (\ref{eq:constraint1}) and
(\ref{eq:constraint2}) double the number of UDD constraints
(\ref{eq:constraint}), and use two different switching functions that
take values 1, $-$1, and 0.

\begin{figure}
  \caption{\label{fig:switching_times} Switching times for orders
    $n=1$ to $n=10$.  UDD switching times are open circles
    \opencircle.  Even permutation $A_3$ DFS DD switching times are
    filled circles \fullcircle only.  Switching times for $S_3$ DFS DD
    over all six $S_3$ permutations include open circles, filled
    circles, and stars $\star$.  The switching times are symmetric
    about $t=\frac{1}{2}$.}
\begin{center}
\includegraphics{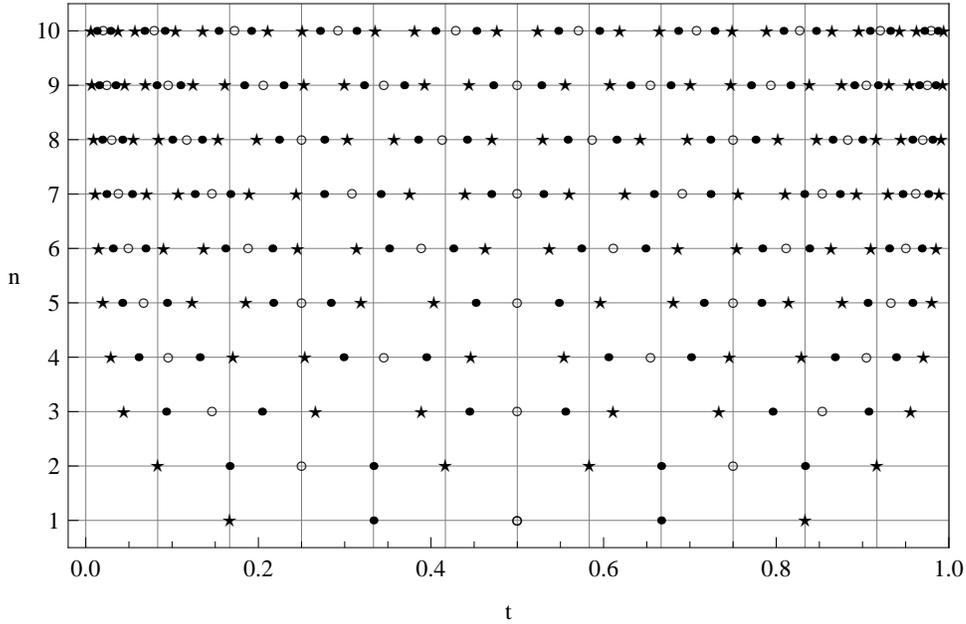}
\end{center}
\end{figure}

We have found numerical solutions to the constraint equations
(\ref{eq:constraint1}) and (\ref{eq:constraint2}) to $64^\mathrm{th}$
order using a basic period four sequence of Hamiltonians
$\{H_1,H_2,H_3,H_2\}$.  Switching from $H_1$ to $H_2$ and from $H_2$
to $H_3$ is accomplished with a $P$ pulse; $P^{-1}$ brings $H_3$ to
$H_2$ and $H_2$ to $H_1$.  The basic period four sequences for the
switching functions are $f_1 = \{1,-1,0,-1\}$ and $f_2 =
\{-1,0,1,0\}$.  For order $n$ decoupling the basic period four
sequence is repeated $\lceil(2n+1)/4\rceil$ times, and the first
$2n+1$ Hamiltonians are taken.  The sequence of pulses $P$ and
$P^{-1}$ is determined from the progression of Hamiltonian types.  Odd
decoupling orders require a final $P$ pulse at the end of the sequence
(in general, the product of the pulses over the whole evolution
interval must be the identity).  For example, third order decoupling
repeats the sequence twice, and takes the first seven entries
$\{H_1,H_2,H_3,H_2,\; H_1,H_2,H_3\}$.  The corresponding pulse
sequence is
$\{\F(\tau_1),P,\F(\tau_2),P,\F(\tau_3),P^{-1},\F(\tau_4),P^{-1},
\F(\tau_5),P,\F(\tau_6),P,\F(\tau_7),P\}$, where $\F(\tau)$ is a free
evolution interval of length $\tau$.  The time intervals $\tau_j$ are
given by the difference $t_j-t_{j-1}$ between successive switching
times.  Figure \ref{fig:switching_times} shows the $A_3$ DFS DD
switching times (filled circles \fullcircle) up to $10^\mathrm{th}$
order satisfying (\ref{eq:constraint1}) and (\ref{eq:constraint2}).
The UDD switching times (open circles \opencircle) are also shown; a
pair of $A_3$ DFS DD switching times brackets each of the UDD
switching times.  Though this bracketing structure is suggestive, we
have been unable to determine a simple analytical formula for the DFS
DD switching times.  Numerical values for these switching times are
given in table \ref{tab:switching_times}.

\begin{table}
  \caption{\label{tab:switching_times} $A_3$ DFS DD switching times
    corresponding to the filled circles in figure \ref{fig:switching_times}.
    Switching times listed are for the interval $[0,1]$.  Only switching times
    less than $\frac{1}{2}$ are given explicitly.  Switching times
    greater than $\frac{1}{2}$ are obtained by reflecting the given
    values about $t=\frac{1}{2}$.} 
\begin{indented}
%\begin{center}
\item[]\begin{tabular}{@{}cl}
\br
order $n$ & $A_3$ switching times \\
\mr
1 & 0.3333333333333333 \\ \mr
2 & 0.1666666666666667 \\  & 0.3333333333333333 \\\mr
3 & 0.0930802599812912 \\  & 0.2041913710924023 \\  & 0.4444444444444444 \\\mr
4 & 0.0611678063574247 \\  & 0.1320291453900112 \\  & 0.2986958120566778 \\  & 0.3945011396907580 \\\mr
5 & 0.0422244245173296 \\  & 0.0940587956886883 \\  & 0.2172228408817372 \\  & 0.2838895075484039 \\  & 0.4518343711713587 \\\mr
6 & 0.0313685011617312 \\  & 0.0691609286752199 \\  & 0.1617103538537611 \\  & 0.2161866929592387 \\  & 0.3514848584641742 \\  & 0.4258827585118745 \\\mr
7 & 0.0239219438795333 \\  & 0.0535688803938237 \\  & 0.1262566342290569 \\  & 0.1675244212375237 \\  & 0.2761133079137736 \\  & 0.3417044666375784 \\  & 0.4698392155798953 \\
\br
\end{tabular}
\;\;
\begin{tabular}{@{}cl}
\br
order $n$ & $A_3$ switching times \\
\mr
8 & 0.0190156712850090 \\  & 0.0422945303794296 \\  & 0.1002297726086257 \\  & 0.1346268067223472 \\  & 0.2239571558152790 \\  & 0.2763447583052162 \\  & 0.3850761827426867 \\  & 0.4416793537112741 \\\mr
9 & 0.0153608717513108 \\  & 0.0344809416081787 \\  & 0.0820319124861268 \\  & 0.1096019013513601 \\  & 0.1835330371665574 \\  & 0.2291713148467980 \\  & 0.3223652733904291 \\  & 0.3688693585992699 \\  & 0.4721657549445159 \\\mr
10 & 0.0127428989292003 \\   & 0.0284688256262034 \\   & 0.0679240161384205 \\   & 0.0914464121824144 \\   & 0.1538757061482468 \\   & 0.1916101903303824 \\   & 0.2714006848897883 \\   & 0.3135792550438800 \\   & 0.4050737288155140 \\   & 0.4525790184049564 \\
\br
\\ \\ \\
\end{tabular}
\end{indented}
%\end{center}
\end{table}

\begin{figure}
  \caption{\label{fig:filter_functions} Filter functions for various
    decoupling orders.  Dashed lines are the UDD filter functions.
    Black and gray lines are the $A_3$ DFS DD filter functions
    corresponding to $f_1^{(n)}(t)$ and $f_2^{(n)}(t)$, respectively.  The
    sets of lines are labeled by their decoupling order $n$.  $n=0$ is
    free evolution.}
\begin{center}
\includegraphics{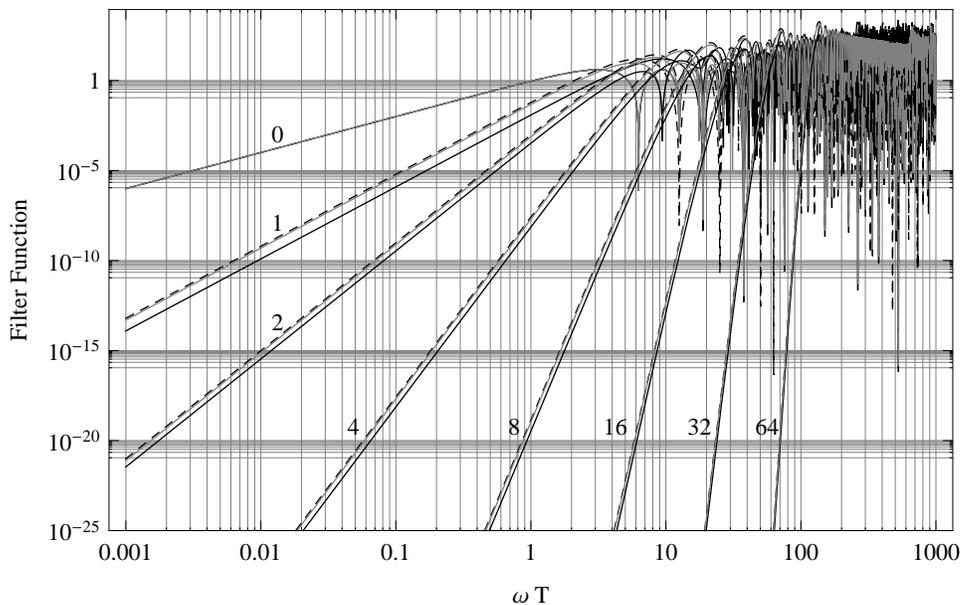}
\end{center}
\end{figure}

Filter functions corresponding to the Fourier transform of the two
order $n$ $A_3$ switching functions $f_1^{(n)}(t)$ and $f_2^{(n)}(t)$ can
be computed analogously to (\ref{eq:filter_function}).  Figure
\ref{fig:filter_functions} shows the filter functions corresponding to
the UDD switching function and the $A_3$ DFS DD switching functions
for various orders of decoupling.  Since all the switching functions
of a given order are designed to integrate the same number of
monomials to zero, they all exhibit the same low frequency filtering
behavior.

We have thus far restricted ourselves to decoupling over even
permutations $A_3$ only.  The previous considerations can be
generalized to decoupling over all permutations of $S_3$.  Defining
the odd permutation Hamiltonian types as
\begin{eqnarray}
H_4 = Z_2 B_1 + Z_1 B_2 + Z_3 B_3, \label{eq:H_4}\\
H_5 = Z_3 B_1 + Z_2 B_2 + Z_1 B_3, \label{eq:H_5}\\
H_6 = Z_1 B_1 + Z_3 B_2 + Z_2 B_3, \label{eq:H_6}
\end{eqnarray}
and computing the accumulated phase difference integrals
(cf. (\ref{eq:diff1}) -- (\ref{eq:diff2})) give rise to five sets of
constraint equations (cf. (\ref{eq:constraint1}) and
(\ref{eq:constraint2})) and five independent $S_3$ switching
functions.  We find that a basic period 10 sequence of
$\{H_1,H_4,H_2,H_5,H_3,H_6,H_3,H_5,H_2,H_4\}$ allows order $n$
decoupling with $5n+1$ evolution intervals.  The period 10 sequence
alternates between even and odd permutations, with the transformation
between successive Hamiltonians accomplished with a single swap gate
on qubits 1 and 2, or qubits 2 and 3.  The corresponding basic
sequence of swap gates is $\{\Po,\Pt,\Po,\Pt,\Po,\Po,\Pt,\Po,\Pt\}$.
As before, the final pulse in the sequence is chosen so that the
pulses multiply to the identity.  We can again solve the constraint
equations numerically, and the switching times for $S_3$ DFS DD are
also displayed in figure \ref{fig:switching_times} (open circles
\opencircle, filled circles \fullcircle, and stars $\star$).  The
switching times for $S_3$ DFS DD are not unique.  Because decoupling
can be accomplished with even permutations alone, or with odd
permutations alone, the relative weighting between even permutation
decoupling and odd permutation decoupling is unconstrained.  The
switching times for $S_3$ DFS DD shown in figure
\ref{fig:switching_times} correspond to equal total time in even and
odd permutations.  Filter functions corresponding to the five $S_3$
switching functions may be computed according to
(\ref{eq:filter_function}).  They have the same low-frequency roll-off
behavior as the UDD filter function.

\begin{table}
  \caption{\label{tab:switching_times_S3} $S_3$ DFS DD switching times
    corresponding to the stars in figure \ref{fig:switching_times}.
    Again, only switching times less than $\frac{1}{2}$ are given explicitly (see the
    caption to table \ref{tab:switching_times}).  Full $S_3$ switching
    times include the values below, the values in table
    \ref{tab:switching_times}, and the UDD values
    (\ref{eq:UDD_switching_times}).}  
\begin{indented}
%\begin{center}
\item[]\begin{tabular}{@{}cl}
\br
order $n$ & $S_3$ switching times \\
\mr
1 & 0.1666666666666667\\\mr
2 & 0.0833333333333333\\ & 0.4166666666666667\\\mr
3 & 0.0441757320558095\\ & 0.2663979542780318\\ & 0.3888888888888889\\\mr
4 & 0.0292438385042891\\ & 0.1706622892054447\\ & 0.2539956225387781\\ & 0.4459105051709558\\\mr
5 & 0.0198486448526978\\ & 0.1234090460471676\\ & 0.1855655208780249\\ & 0.3188988542113582\\ & 0.4035604011944698\\\mr
6 & 0.0148169093658703\\ & 0.0902375649702305\\ & 0.1365285435153074\\ & 0.2455460335064556\\ & 0.3140624574739186\\ & 0.4629576452117436\\\mr
7 & 0.0112075501170748\\ & 0.0704161635276064\\ & 0.1069644388480345\\ & 0.1894875256294779\\ & 0.2440254284265229\\ & 0.3752930054921819\\ & 0.4396659439243005\\
\br
\end{tabular}
\;\;
\begin{tabular}{@{}cl}
\br
order $n$ & $S_3$ switching times \\
\mr
8 & 0.0089377851765520\\ & 0.0553969487226799\\ & 0.0843798400197465\\ & 0.1531864662990289\\ & 0.1981928332289538\\ & 0.3030094142447375\\ & 0.3573598562810042\\ & 0.4706108187731435\\\mr
9 & 0.0071855182206674\\ & 0.0453635718581324\\ & 0.0692317973728011\\ & 0.1243583767250380\\ & 0.1614266670585074\\ & 0.2527437110834672\\ & 0.2994843851270477\\ & 0.3925059539229590\\ & 0.4443099124772396\\\mr
10 & 0.0059745260011464\\ & 0.0373611383696360\\ & 0.0571010082270104\\ & 0.1041446643790700\\ & 0.1355161859798696\\ & 0.2110062905707306\\ & 0.2508943643743651\\ & 0.3352721033248206\\ & 0.3812593890562080\\ & 0.4762946103276755\\
\br
\\ \\ \\
\end{tabular}
\end{indented}
%\end{center}
\end{table}

\section{Decoupling the 3-qubit DFS from a quantum bath}
\label{sec:quantum_dd}
A DFS qubit coupled to local quantum baths is described by the
Hamiltonian 
\begin{equation}
H = B_0 + \vec{S}_1.\vec{B}_1 + \vec{S}_2.\vec{B}_2 + \vec{S}_3.\vec{B}_3.
\label{eq:qH}
\end{equation}
Here $B_0$ is a pure bath operator describing interactions within the
bath alone, and $\vec{B}_j=(B_{j,x},B_{j,y},B_{j,z})$ is a vector of
bath operators coupled to constituent qubit $j$ through its Pauli
operators $\vec{S}_j=(X_j,Y_j,Z_j)$.  As in section
\ref{sec:classical_dd} full swaps are used to permute the qubits
through the different bath operators.  We define the quantum bath
Hamiltonians analogously to (\ref{eq:H_1}),
(\ref{eq:H_2})--(\ref{eq:H_3}), and (\ref{eq:H_4})--(\ref{eq:H_6}):
\begin{eqnarray}
H_1 = B_0 + \vec{S}_1.\vec{B}_1 + \vec{S}_2.\vec{B}_2 + \vec{S}_3.\vec{B}_3,\label{eq:H_1q}\\
H_2 = B_0 + \vec{S}_2.\vec{B}_1 + \vec{S}_3.\vec{B}_2 + \vec{S}_1.\vec{B}_3,\label{eq:H_2q}\\
H_3 = B_0 + \vec{S}_3.\vec{B}_1 + \vec{S}_1.\vec{B}_2 + \vec{S}_2.\vec{B}_3,\label{eq:H_3q}\\
H_4 = B_0 + \vec{S}_2.\vec{B}_1 + \vec{S}_1.\vec{B}_2 + \vec{S}_3.\vec{B}_3,\label{eq:H_4q}\\
H_5 = B_0 + \vec{S}_3.\vec{B}_1 + \vec{S}_2.\vec{B}_2 + \vec{S}_1.\vec{B}_3,\label{eq:H_5q}\\
H_6 = B_0 + \vec{S}_1.\vec{B}_1 + \vec{S}_3.\vec{B}_2 + \vec{S}_2.\vec{B}_3. \label{eq:H_6q}
\end{eqnarray}

The first and second order $A_3$ and $S_3$ DFS DD sequences for
decoupling classical phase noise also decouple the DFS qubit from
quantum baths.  Consider the first order $A_3$ evolution sequence
\begin{eqnarray}
\lefteqn{P.\exp(-\rmi H \tau).P. \exp(-\rmi H \tau).P. \exp(-\rmi H \tau)}\\
 &=& \exp(-\rmi H_3 \tau).\exp(-\rmi H_2 \tau).\exp(-\rmi H_1 \tau)\label{eq:qdd1}\\
&\approx&1-\rmi\tau \left(3B_0 + \vec{S}_\mathrm{tot}.\vec{B}_\mathrm{tot}\right),
\end{eqnarray}
where 
\begin{eqnarray}
\vec{S}_\mathrm{tot}\equiv \vec{S}_1+\vec{S}_2+\vec{S}_3,\label{eq:Stot}\\
\vec{B}_\mathrm{tot}\equiv \vec{B}_1+\vec{B}_2+\vec{B}_3.\label{eq:Btot}
\end{eqnarray}
The first order effective Hamiltonian contains only coupling to the
total spin component operators of all three qubits.  Similarly, the
second order $A_3$ evolution sequence is
\begin{eqnarray}
  \hspace{-0.5in} \lefteqn{\exp(-\rmi H \tau).P^{-1}.\exp(-\rmi H \tau).P^{-1}.\exp(-2\rmi H \tau).P. \exp(-\rmi H \tau).P. \exp(-\rmi H \tau)}\\
  &=& \exp(-\rmi H_1 \tau).\exp(-\rmi H_2 \tau).\exp(-2\rmi H_3 \tau).\exp(-\rmi H_2 \tau).\exp(-\rmi H_1 \tau)\label{eq:qdd2}\\
  &\approx&1-2\rmi\tau \left(3B_0 + \vec{S}_\mathrm{tot}.\vec{B}_\mathrm{tot}\right)
  -2\tau^2 \left(3B_0 + \vec{S}_\mathrm{tot}.\vec{B}_\mathrm{tot}\right)^2.
\end{eqnarray}
Again, the effective Hamiltonian depends to second order only on the
total spin component operators.  The first and second order $S_3$ DFS
DD sequences can be expanded similarly and also decouple quantum
baths.

For third and higher order, both the $A_3$ and $S_3$ sequences fail to
decouple the quantum bath to the nominal order.  To find higher order
sequences for decoupling the quantum bath, we have resorted to brute
force computational searches.  We seek a sequence of Hamiltonians
$H_{\sigma(k)}$ and associated time intervals $\tau_k$ such that the
product of unitary propagators
\begin{equation}
  U = \prod_{k=1}^N \exp(-\rmi H_{\sigma(k)} \tau_k)
\label{eq:Uproduct}
\end{equation}
has an effective Hamiltonian that couples only to the total spin
operators $S_{\mathrm{tot},x}$, $S_{\mathrm{tot},y}$, and
$S_{\mathrm{tot},z}$, to the desired decoupling order.  Here $N$ is the
total number of evolution intervals and $\sigma(k)$ specifies under
which Hamiltonian (\ref{eq:H_1q})--(\ref{eq:H_6q}) interval $k$
evolves.  

Given some candidate sequence of Hamiltonians, we expand
(\ref{eq:Uproduct}) to the desired decoupling order in $\tau$.
Because there are three qubits in the DFS, the system-bath interaction
terms are at most weight three Pauli products on the system
(constituent) qubits.  The set of 64 Pauli products of weight three or
less is partitioned into 16 equivalence classes under the action of
the group $S_3$.  For example, $X_1$ is in the equivalence class
consisting of $[X_1,X_2,X_3]$, and $X_1 Y_2$ is in the equivalence
class consisting of $[X_1Y_2, X_1Y_3, Y_1X_2, Y_1X_3, X_2Y_3,
Y_2X_3]$.  In the order-by-order expansion of (\ref{eq:Uproduct}) the
coefficients of the Pauli products in the same equivalence class must
be equal in order to achieve an effective Hamiltonian that couples to
global interactions alone.  Note that the coefficients consist of
products of bath operators, with some numerical prefactor.  Because
the various bath operators do not commute, they further grade the
terms that must be set equal.  A candidate Hamiltonian sequence is a
valid decoupling solution if $\tau_k$'s can be found such that members
of each of the equivalence classes have the same coefficients, with the
$\tau_k$'s real and positive.  For the examples above, if the
members of the $[X_1]_{S_3}$ equivalence class all couple to the bath
operator $C$, the corresponding global interaction term is
$S_{\mathrm{tot},x}.C$.  Similarly, if the members of the
$[X_1Y_2]_{S_3}$ equivalence class all couple to the bath operator
$D$, the corresponding global interaction terms are
$(S_{\mathrm{tot},x}S_{\mathrm{tot},y}-\rmi S_{\mathrm{tot},z}).D$.

%Mathematica reference?

With this methodology, we have found the third order quantum bath
decoupling sequence shown in table \ref{tab:qdd3}.  Rather than giving
switching times, we have listed the evolution time intervals
corresponding to a total evolution time of 1.  The sequence consists
of 26 Hamiltonian intervals, with a ``doubly palindromic'' structure.
The first 13 Hamiltonians are all even permutations while the second
13 are all odd permutations.  The second 13 Hamiltonian types can be
determined from the first 13 via the mapping $H_1\rightarrow H_4$,
$H_2\rightarrow H_6$, and $H_3\rightarrow H_5$.  The interval timings
for the odd permutations are identical to the even permutations.
Within each set of 13 intervals, the interval timings are palindromic.
Other length 26 third order solutions have been found, but the one
displayed in table \ref{tab:qdd3} is optimal in the sense that the
ratio between the maximum and minimum interval lengths is smallest.

\begin{table}
  \caption{\label{tab:qdd3} Third order DFS decoupling sequence for a
    quantum bath, consisting of 26 Hamiltonian intervals.  The first
    column lists the Hamiltonian type
    (\ref{eq:H_1q})--(\ref{eq:H_6q}).  The second
    column gives the interval length (rather than the switching times).  The
    third column gives the pulse required at the end of the Hamiltonian
    interval.  Only the first 13 intervals are given; the interval timings
    and pulses for the second 13 intervals are identical.}
\begin{indented}
\item[]\begin{tabular}{@{}ccl}
\br
Hamiltonian type & interval length & pulse \\
\mr
$H_1$ & 0.02443154605193963 & $P$      \\
$H_2$ & 0.03273388118971666 & $P$      \\
$H_3$ & 0.05269740572865081 & $P^{-1}$ \\
$H_2$ & 0.03073701555573789 & $P^{-1}$ \\
$H_1$ & 0.04633548169315730 & $P^{-1}$ \\
$H_3$ & 0.05049836419256131 & $P$      \\
$H_1$ & 0.02513261117647280 & $P$      \\
$H_2$ & 0.05049836419256131 & $P^{-1}$ \\
$H_1$ & 0.04633548169315730 & $P^{-1}$ \\
$H_3$ & 0.03073701555573789 & $P^{-1}$ \\
$H_2$ & 0.05269740572865081 & $P$      \\
$H_3$ & 0.03273388118971666 & $P$      \\
$H_1$ & 0.02443154605193963 & $\Po$    \\
\br
\end{tabular}
\end{indented}
\end{table}

\section{Numerical Simulations}
\label{sec:simulations}

\begin{figure}
  \caption{\label{fig:classical_bath} Simulations of a DFS qubit
    coupled to a classical dephasing-only bath.  Infidelity is plotted
    against total evolution time $T$ for $A_3$ DFS DD orders $n=0$
    (free evolution) through $n=4$.  $n=0$ infidelities are given by
    \fullcircle, $n=1$ by \fullsquare, $n=2$ by $\blacklozenge$, $n=3$
    by $\blacktriangle$, $n=4$ by $\blacktriangledown$.  Dotted lines
    are fits to the given leading order infidelity term.}
\begin{center}
\includegraphics{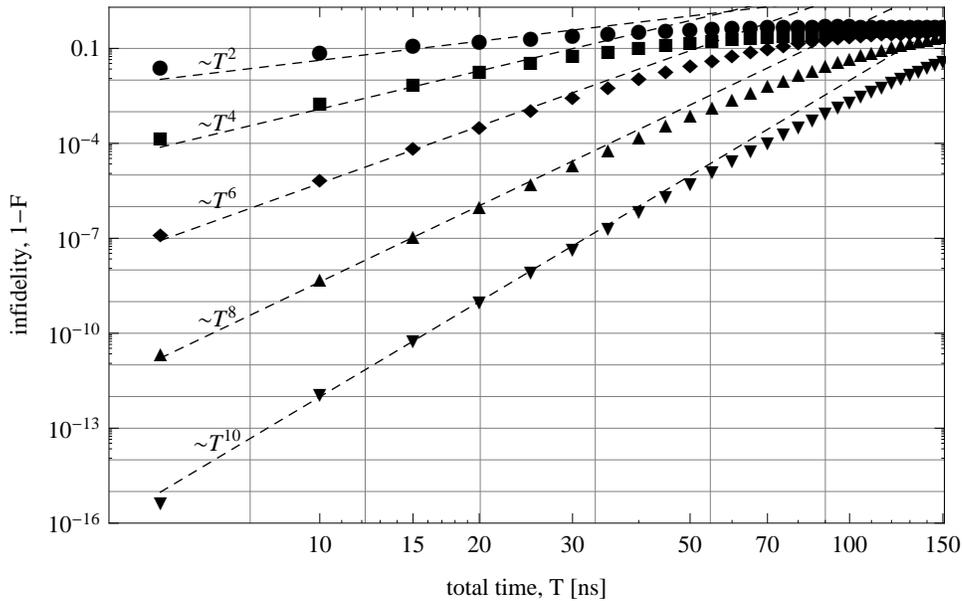}
\end{center}
\end{figure}

Numerical simulations of a DFS qubit coupled to classical and quantum
baths have been performed to illustrate the performance of the DFS
decoupling sequences.  Figure \ref{fig:classical_bath} shows a log-log
plot of the infidelity versus total evolution time for a DFS qubit
coupled to a classical dephasing only bath, with $A_3$ DFS DD
sequences of orders $n=0$ to $n=4$ applied to the DFS qubit.  The
infidelity is defined as $1-F$, where $F$ is given by
(\ref{eq:fidelity3}).  The classical bath functions have size
$|B_j(t)|\sim100$MHz and $|(dB_j/dt)/B_j|\sim100$MHz.  Each point in
figure \ref{fig:classical_bath} gives the infidelity after a single
round of $n^\mathrm{th}$ order decoupling for the corresponding total
evolution time $T$ on the abscissa; each point is averaged over 50
dephasing bath instances and 100 initial encoded DFS states.  The plot
markers correspond to $A_3$ DFS DD orders $n=0$ through $n=4$ as
described in the figure caption. For $n^\mathrm{th}$ order decoupling,
the phase differences (\ref{eq:diff1})--(\ref{eq:diff3}) at short
total times $T$ have dependence $\sim T^{n+1}$; the corresponding
infidelity has total time dependence $\sim T^{2(n+1)}$.  The dotted
lines in figure \ref{fig:classical_bath} are fits to the function $a-b
T^{2(n+1)}$ for each order of decoupling.  At short total times the
infidelity scales as expected, with a straight-line dependence on the
log-log plot.  The slopes of the fits have the predicted values of
$2(n+1)$, indicating that the DFS DD sequences decouple the dephasing
baths to their designed orders.

\begin{figure}
  \caption{\label{fig:quantum_bath} Simulations of a DFS qubit
    coupled to a quantum bath.  Infidelity is plotted
    against total evolution time $T$ for DFS DD orders $n=0$
    (free evolution) through $n=3$. $n=1$ and $n=2$ use $S_3$ DFS DD
    pulse sequences; $n=3$ uses the pulse sequence in table \ref{tab:qdd3}.
    $n=0$ infidelities are given by
    \fullcircle, $n=1$ by \fullsquare, $n=2$ by $\blacklozenge$, $n=3$
    by $\blacktriangle$.  Dotted lines
    are fits to the given leading order infidelity term.}
\begin{center}
\includegraphics{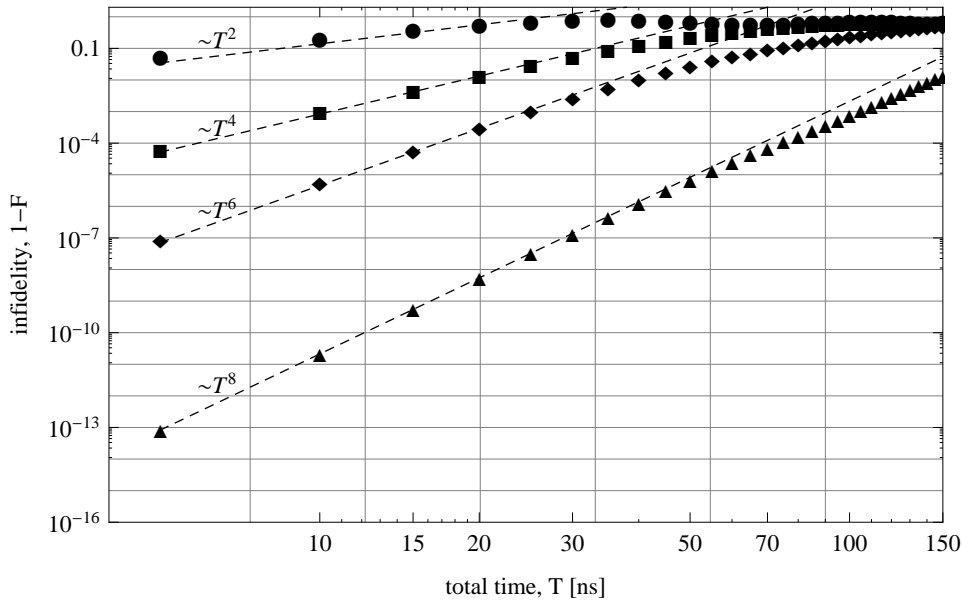}
\end{center}
\end{figure}

Figure \ref{fig:quantum_bath} shows a log-log plot of infidelity
versus total evolution time for a DFS qubit coupled to a quantum bath.
We simulate a DFS qubit coupled to a bath composed of six spins; each
constituent system qubit is coupled to two bath spins, and all six
bath spins are coupled to each other.  The Hamiltonian simulated is
\begin{equation}
\hspace{-0.5in}H=J \left(
\sum_{j=1}^2 r_{1,j}\vec{S}_1.\vec{I}_j
+\sum_{j=3}^4 r_{2,j}\vec{S}_2.\vec{I}_j
+\sum_{j=5}^6 r_{3,j}\vec{S}_3.\vec{I}_j
\right)
+ \beta \sum_{j=1}^6\sum_{k=j+1}^6 r_{j,k} \vec{I}_j.\vec{I}_k,
\label{eq:simulated_hamiltonian}
\end{equation}
which has the same structure as (\ref{eq:qH}).  In
(\ref{eq:simulated_hamiltonian}) $J$ gives the energy scale of the
system-bath coupling, $\beta$ gives the energy scale of the intra-bath
coupling, $r_{i,j}$ are random numbers between 0 and 1, $\vec{S}_j$ is
the vector of Pauli operators for the $j^\mathrm{th}$ system spin, and
$\vec{I}_j$ is the vector of Pauli operators for the $j^\mathrm{th}$
bath spin.  We take $J=100$MHz and $\beta=$10KHz for the simulations
in figure \ref{fig:quantum_bath}.  The simulations are ``numerically
exact'' in that the propagators for full system and bath evolution are
determined by computing the matrix exponential of the Hamiltonian
(\ref{eq:simulated_hamiltonian}); swaps between system qubits are
interspersed between the propagators as required by the DFS DD
protocols.  Fidelities are again computed according to
(\ref{eq:fidelity3}), with the states, propagator $U$, and projector
$\Pi$ suitably generalized to include both the system and bath, and
the addition of a partial trace operation over the bath spins.  Each
data point in figure \ref{fig:quantum_bath} shows the infidelity
computed for a given total evolution time, for decoupling orders $n=0$
through $n=3$.  Each data point gives the infidelity averaged over 52
random initial conditions and Hamiltonians.  The first and second
order decoupling sequences used are $S_3$ DFS DD sequences; the third
order sequence used is given in table \ref{tab:qdd3}.  As in the
classical bath case, the infidelity is expected to scale as
$T^{2(n+1)}$.  The dotted lines giving fits to the function $a-b
T^{2(n+1)}$ confirm the expected scaling.

\section{Conclusion}
\label{sec:conclusion}
We have shown that exchange pulses alone are sufficient to decouple
the 3-qubit DFS from classical and quantum baths.  By averaging over
permutations of the 3 constituent qubits, local baths can be made to
appear global to high order.  Because the 3-qubit DFS is immune to
global decoherence, DFS DD protects the encoded information.
Numerical simulations of the new DFS decoupling pulse sequences
confirm DD sequence performance expected from analytical
considerations.

Decoupling of the 3-qubit DFS from classical and quantum baths may
also be accomplished using NUDD (nested Uhrig dynamical decoupling)
pulse sequences \cite{Wang:2011dq}.  The DFS DD pulse sequences,
however, are far more efficient.  For example, for third order
decoupling from a quantum bath, $n=3$, NUDD requires
$(n+1)^{2\times3}=4096$ pulse intervals, compared to 26 pulse
intervals for the DFS decoupling described in section
\ref{sec:quantum_dd}.  Accounting for the structure of the
DFS---protecting the encoded information only and using exchange
pulses only---substantially reduces the number of pulses needed for
decoupling and removes the need for single qubit Pauli gates.  The
fact that only a particular subspace or subsystem needs to be
protected should be used in designing efficient decoupling pulse
sequences for general qubit encodings.

Decoupling by averaging over the symmetric group (or one of its
subgroups) using exchange pulses can be generalized to other
decoherence free subspaces and subsystems \cite{Kempe:2001a}.  The
most efficient sequences as well as the types of errors protected
against will depend on the encoding, the structure of the system-bath
coupling, and the noise model.  For the 2-qubit decoherence free
subspace, for example, in which the encoding protects against
collective decoherence in a single direction (say $z$), decoupling
over $S_2$ protects against bath variations in the $z$ direction only.
Bath variations in the $x$ and $y$ directions will cause leakage out
of the encoded subspace.  Decoupling over $S_2$ has exactly the same
structure as UDD for a single qubit: a full swap on the exchange gate
between the two qubits takes the role of the Pauli $X$ pulse for the
single qubit, and the interval timings are the UDD timings
(\ref{eq:UDD_switching_times}).  For other encodings protecting
against weak collective decoherence \cite{Kempe:2001a}, averaging over
$S_n$ subgroups will decouple bath variations in the
encoding-protected direction only.  For encodings protecting against
strong collective decoherence (which includes the 3-qubit DFS)
averaging over $S_n$ subgroups decouples the encoded information from
all bath components.  The effective global interaction created by the
averaging affects only the gauge while leaving the encoded information
unchanged.  The quantum numbers describing encoded states correspond
to total spin operators on increasing numbers of the constituent
qubits, and all total spin operators commute with global interactions.
Determination of switching times and pulse interval Hamiltonian types
for decoupling other DFS encodings can be found by generalizing the
methods described in this paper.  

Finally, we note that the correspondence between $S_2$ decoupling and
UDD of a 2-level system, each coupled to classical dephasing bath(s)
along a single direction, generalizes to a correspondence between
decoupling a $d$-qubit DFS and a $d$-level system ($SU(d)$) from
classical dephasing baths along a single direction
\cite{Shukla:2012qy}.  In these cases averaging over the $d$ cyclic
permutations of the system for the DFS or cyclic permutations of the
$d$ levels yields the same pulse interval Hamiltonian types and
switching times.

\ack
Sponsored by United States Department of Defense.  The views and
conclusions contained in this document are those of the authors and
should not be interpreted as representing the official policies,
either expressly or implied, of the United States Department of
Defense or the U.S. Government.  
Approved for public release, distribution unlimited.
% Approved March 6, 2012

\appendix
\section*{Appendix}
\label{sec:fidelity}
\setcounter{section}{1}
For an initial encoded state $\ket{\psi_e}=(r,\sqrt{1-r^2}e^{\rmi
  \phi})$, the coefficients in the fidelity expression
(\ref{eq:fidelity3_explicit}) are
\begin{eqnarray}
c_0&=&\frac{1}{6} \left(3-2r^2+2r^4+2r^2(1-r^2)\cos2\phi\right),\\
c_1&=&\frac{2}{9}(1-r^2)\left(1+2r^2+2r\sqrt{3(1-r^2)}\cos\phi\right),\\
c_2&=&\frac{2}{9}(1-r^2)\left(1+2r^2-2r\sqrt{3(1-r^2)}\cos\phi\right),\\
c_3&=&\frac{1}{18}\left(1-2r^2+10r^4-6r^2(1-r^2)\cos2\phi\right).
\end{eqnarray}

\section*{References}
%\bibliography{Quantum_Computing.bib}
%\bibliographystyle{unsrt}

\end{document}